\documentclass[12pt,fleqn]{article}
\usepackage{amsmath}
\usepackage{graphicx}
\usepackage{multirow}

\begin{document}
\bibliographystyle{prsty}
\title{Comparison of variational and CSE methods to  polaron ground-state energy}
\author{ Qing Chen$^{1}$,  Ke-Lin Wang$^{1}$, Yi Wang$^{2}$
  \\
{\small\it $^{1}$Department of Modern Physics,$^{2}$Department of Physics}\\
{\small\it University of Science and Technology of China, Hefei
230026, PR China}} \maketitle

\begin{abstract}
\baselineskip20pt
    Taking the same trial wave function, the ground-state energy of Fr\"{o}hlich polaron is investigated
    by variational method and coherent-state expansion (CSE) one, respectively.
    Within the accuracy to $\alpha^{2}$(the electron-phonon coupling constant), both methods can give the
    same analytic expressions of polaron ground-state energy as the function of electron-phonon coupling constant.
    We find that the CSE method can much simplify the calculation and shows more advantages in higher order approximations.

\end{abstract}

\newpage

\baselineskip20pt
Electron-phonon (e-p) interaction is an important elementary excitation in condensed matter
physics. Although the Green Function method is quite mature to deal with such problems, 
it still can't provide ideal results when the e-p interaction is relatively
strong, such that, the variational method were often involved in consideration. Polaron problem is particular notable in 
ionic or semiconductor materials. Various approaches $^{{\cite{Frolich}}-{\cite{review2}}}$ had been applied to the calculation of 
polaron ground-state energy during the past 40 years. 
Recently, based on an effective Hamiltonian of Fr\"{o}hlich polaron under LLP transformation,
a new concise method $^{\cite{kelin}}$ was proposed by Wang et al. by expanding the wave function
of phonon's coherent-state in powers of phonon's correlation. Putting the wave function into Sch\"{o}dinger equation,
${\bf H}|>=E|>$
and comparing the coefficients of same power of phonon's creation operator in two sides of the equation,
a set of coupled equations can be obtained, and rather good results of ground-state energy and system wave function 
can be calculated by iteration method, especially in a wide range of coupling strength materials.
Differently from variational method and other methods, coherent-state expansion (CSE) is systematic, concise and effective
exhibited in two aspects: the first is that the establishment of the wave function will be not changed  
 in different problem, the second is that we can expand the wave function to higher order phonon's correlation. 
In fact, the great success of CSE method have been shown in the study of mobile polaron$^{\cite{yi}}$, 
exciton-phonon system$^{\cite{exciton}}$ and dissipative two-level system$^{\cite{Jiang}}$etc.

The systematic CSE study with higher order phonon's correlation had been performed in 1D polaron$^{\cite{1DP}}$ system
and other systems $\cite{Jiang}$,$\cite{Han}$,${\cite{Holstein}}$. As an example, 
on the physical study of two-site Holstein model$^{\cite{Holstein}}$, CSE calculations are relatively simple
compared to other approaches. Theoretically, CSE trial wave function can be expanded to arbitrary orders,  
it is just an exact example to demonstrate the "convergence" of CSE method. In order to illustrate the 
reliability and applicability of CSE method and provide a new choice to other physicists, we adopt the same wave 
function as the article${\cite{kelin}}$ and use variational method to study the intermediate coupling polaron 
ground-state energy. In 1968 Larsen had studied the polaron ground state energy and effective mass by using 
variational method with a similar variational wave function$^{\cite{Larsen}}$.
He  studied the static polaron problem and the wave function he took is a
special case of CSE method. In addition, the second-order explicit expression of polaron ground-state energy
as a function of electron-phonon coupling constant had not been given in previous papers ${\cite{Larsen}}$,
${\cite{kelin}}$ and ${\cite{yi}}$. In present paper we will discuss and compare the calculation results
under these two methods.

By adopting the units of $2m=\omega_{0} =V=h=1$, the Fr\"{o}hlich polaron Hamiltonian reads
\begin{equation}
\label{eq:1.1}
{\bf H}={\bf P}^{2}+\sum\limits_{{\bf q}}a_{{\bf q}}^{+}a_{{\bf q}}+
\sum\limits_{{\bf {\bf q}}}V_{{\bf q}}(a_{{\bf q}}e^{i{\bf q \cdot r}}+a_{{\bf q}}^{+}e^{-i{\bf q \cdot r}}),
\end{equation}
where
\begin{equation}
\mid{V_{\bf q}}\mid^2=\left
\{\frac{\Gamma(\frac{N-1}{2})2^{N-3/2}\pi^{(N-1)/2}}{V_N{\bf
q}^{N-1}}\right \}\alpha ,
\end{equation}
$a_{{\bf q}}^{+}$ and $a_{{\bf q}}$ are creation and annihilation operators of
the longitudinal optical(LO) phonon with wave vector ${\bf q}$, respectively.
$\alpha$ is a dimensionless electron-phonon coupling constant, N is the space dimensionality.
After proceeding of LLP$^{\cite{LLP}}$ unitary transformation, the polaron 
Hamiltonian becomes
\begin{equation}
\label{eq:LLPH}
H=\sum_{{\bf q}}(1-2{\bf Q \cdot q}+{\bf q}^2)a_{{\bf q}}^{\dag}a_{{\bf q}}+
 \sum_{{\bf q,q'}}
{\bf q \cdot q'}a_{{\bf q}}^{\dag}a_{{\bf q'}}^{\dag}a_{{\bf q}}a_{{\bf q'}}
+\sum_{{\bf q}}V_{{\bf q}}(a_{{\bf q}}^{\dag}+a_{{\bf q}}).
\end{equation}
where ${\bf Q}$ is total momentum of a polaron, 
${\bf Q}={\bf P}+\sum\limits_{{\bf q}}{\bf q}a_{{\bf q}}^{+}a_{{\bf q}}$, and it is a 
conserved quantity. In order to compare with CSE method, we assume the variational wave function is the same as Refs. [6][7]
\begin{equation}
\label{eq:3.05}
|>=|>_{0}+\sum\limits_{{\bf q},{\bf q}'}b_{\bf q,q'}a_{{\bf q}}^{+}a_{{\bf q}'}^{+}|>_{0}
\end{equation}
where
\begin{equation}
|>_{0}= {\prod_{{\bf q}}}exp[\alpha_{\bf q}a_{{\bf q}}^{\dag}]|0> .
\end{equation}
Based on the eq.\ref{eq:LLPH}, $\alpha_{{\bf q}}$ and $b_{{\bf q},{\bf q}'}$ are supposed to be the 
real variational parameter. The ground-state energy of the system is  
\begin{equation}
\label{eq:EE}
E={\frac{<|H|>}{<|>} }.
\end{equation}

The aim of present paper is to prove the reliability of CSE method from another point of view, 
therefore we only expand the formula \ref{eq:EE} into a square power of e-p coupling constant and the terms with
higher order $\alpha$ are neglected during the derivation. One finally has
\begin{eqnarray}
\label{eq:E2}
E&=& \sum\limits_{{\bf q}}A_{{\bf q}}\alpha_{{\bf q}}^{2}+2V_{{\bf q}}\alpha_{{\bf q}}+\sum\limits_{{\bf q,q'}}{\bf q
\cdot q'}\alpha_{{\bf q}}^{2}\alpha_{{\bf q}'}^{2} \nonumber \\
&+ &2\sum\limits_{{\bf q,q'}}(b_{{\bf q},{\bf q}'}(A_{{\bf q}}+A_{{\bf q}'}+
2{\bf q \cdot q'})\alpha_{{\bf q}}\alpha_{{\bf q}'}+V_{{\bf q}}\alpha_{{\bf q}'}+V_{{\bf q}'}\alpha_{{\bf q}})\nonumber \\
&+ &2\sum\limits_{{\bf q,q'}}(b_{{\bf q},{\bf q}'}^2(A_{{\bf q}}+A_{{\bf q}'}+2{\bf q \cdot q'}))
\end{eqnarray}
where $A_{{\bf q}}=1-2{\bf Q \cdot q}+{\bf q}^2$. By variation energy (\ref{eq:E2}) with respect to $\alpha_{{\bf q}}$ and
$b_{{\bf q},{\bf q}'}$, we get 
\begin{equation}
 A_{{\bf q}}\alpha_{{\bf q}}+V_{{\bf q}}+\sum\limits_{{\bf q'}}\alpha_{{\bf q}'}^2\alpha_{{\bf q}}{\bf q \cdot q'}+
2\sum\limits_{{\bf q'}}b_{{\bf q},{\bf q}'}[(A_{{\bf q}}+A_{{\bf q}'}+2{\bf q \cdot q'}))
\alpha_{{\bf q}'}+V_{{\bf q}'}]=0
\end{equation}
\begin{equation}
(A_{{\bf q}}+A_{{\bf q}'}+
2{\bf q \cdot q'})\alpha_{{\bf q}}\alpha_{{\bf q}'}+V_{{\bf q}}\alpha_{{\bf q}'}+V_{{\bf q}'}\alpha_{{\bf q}}+
2(b_{{\bf q},{\bf q}'}(A_{{\bf q}}+A_{{\bf q}'}+
2{\bf q \cdot q'})=0
\end{equation}
It is easy to get the iterating equations as follows
\begin{equation}
\label{eq:3}
b_{{\bf q},{\bf q}'}=-{\frac{\alpha_{{\bf q}}\alpha_{{\bf q}'}}{2}}-
{\frac{V_{{\bf q}}\alpha_{{\bf q}'}+V_{{\bf q}'}\alpha_{{\bf q}}}{2(A_{{\bf q}}+A_{{\bf q}'}+
2{\bf q \cdot q'})}}
\end{equation}
\begin{equation}
\label{eq:5}
\alpha_{{\bf q}}=-{\frac{V_{{\bf q}}}{A_{{\bf q}}}}+
 {\frac{1}{A_{{\bf q}}}}\sum\limits_{{\bf q}'}\left\{\alpha_{{\bf q}'}^{2}(A_{{\bf q}}+A_{{\bf q}'}+
{\bf q \cdot q'})\alpha_{{\bf q}}+2\alpha_{{\bf q}}\alpha_{{\bf q}'}V_{{\bf q}'}+\alpha_{{\bf q}'}^{2}V_{{\bf q}}+
V_{{\bf q}'}{\frac{V_{{\bf q}}\alpha_{{\bf q}'}+V_{{\bf q}'}\alpha_{{\bf q}}}{A_{{\bf q}}+A_{{\bf q}'}+
2{\bf q \cdot q'}}}\right\}
\end{equation}

Taking the initial value of $\alpha_{{\bf q}} = -V_{{\bf q}}/A_{{\bf q}}$,  substituting it into (\ref{eq:3}) and 
(\ref{eq:5}) to proceed the iteration, and then put it into (\ref{eq:E2}), 
thus the exact expression of ground-state energy as the function of e-p coupling constant $\alpha^{2}$ 
can be derived as following
\begin{equation}
\label{eq:Etotel}
E={\bf Q}^2 -\sum\limits_{{\bf q}}\frac{V_{{\bf q}}^{2}}{A_{{\bf q}}}-
\sum\limits_{{\bf q,q'}}\frac{2{\bf q \cdot q'}V_{{\bf q}}^{2}V_{{\bf q}'}^{2}}
{A_{{\bf q}}^{2}A_{{\bf q}'}(A_{{\bf q}}+A_{{\bf q}'}+2{\bf q \cdot  q}')}.
\end{equation}

The second energy term of N-dimensional polaron reads
\begin{equation}
\label{eq:E(2)}
E^{(2)}=\sum\limits_{{\bf q,q'}}\frac{2{\bf q \cdot q'}V_{{\bf q}}^{2}V_{{\bf q}'}^{2}}{A_{{\bf q}}^{2}A_{{\bf q}'}(A_{{\bf q}}+A_{{\bf q}'}+
2{\bf q \cdot  q}')}
\end{equation}

Taking N=3 and ${\bf Q}=0$, the system ground-state energy can be given by
\begin{equation}
\label{eq:E159}
E=-\alpha-0.0159196\alpha^{2}
\end{equation}

For CSE method$^{\cite{kelin}}$$^{\cite{yi}}$, the ground-state energy 
and coefficient $b_{{\bf q},{\bf q}'}$ are given by
\begin{equation}
\label{eq:cE}
E={\bf Q}^{2}+\sum\limits_{{\bf q}}V_{{\bf q}}\alpha_{{\bf q}},
\end{equation}
\begin{equation}
\label{eq:calpha}
\alpha_{{\bf q}}=-{\frac{V_{{\bf q}}}{A_{{\bf q}}}}+{\frac{1}{A_{{\bf q}}}}\sum\limits_{{\bf q}'}V_{{\bf q}'}
\frac{2{\bf q \cdot q}'\alpha_{{\bf q}}\alpha_{{\bf q}'}}{2+({\bf q}+{\bf q}')^{2}},
\end{equation}
\begin{equation}
\label{eq:cb}
b_{{\bf q},{\bf q}'}=-{\frac{{\bf q \cdot q'}
\alpha_{{\bf q}}\alpha_{{\bf q}'}}{A_{{\bf q}}+A_{{\bf q}'}+
2{\bf q \cdot q'}}}
\end{equation}

In terms of Eqs.(\ref{eq:cE})-(\ref{eq:cb}), the same expressions of ground-state energy (\ref{eq:Etotel})-(\ref{eq:E159})
can be obtained when it is 
taken within the accuracy of $\alpha^{2}$.

From above calculations we see, the variational method gives the same analytic expression of polaron ground-state energy 
with CSE method in the accuracy of $\alpha^{2}$. This energy formula had been proved correct
by comparing it to the exact numerical value in \cite{kelin}. Because the variational method is a very mature technique, 
our calculation may help us to understand the CSE method better in another point of view.

As an example, if the trial wave function is taken in a third-order approximations, 
the variational derivation will show more complex than CSE method and  yield a tedious mutually coupled equations. 
But for CSE method, it can easily get the third-order(see (\ref{eq:cE}),(\ref{eq:calpha}),(\ref{eq:cb}))
 ground state energy$^{\cite{kelin}}$, which can be compared nicely with 
the sixth perturbation method results by Smondyrev$^{\cite{Smondyrev}}$ and Selyuin$^{\cite{Selyugin}}$.  
It can be concluded that the CSE is a good approximate method.

We should emphasize that, although the same analytic expressions of ground state energy 
can be obtained in the same accuracy of $\alpha^{2}$ by two methods, they are intrinsically different in the 
calculation of higher order approximations.
Comparatively, CSE method has a more precise form and a much simple calculation.
We believe that CSE method will show a potential applications in other analogous systems.


\begin{thebibliography}{100}

\bibitem{Frolich} Frolich H 1954 {\it Philos.Mag.Suppl. } {\bf 3} 325
\bibitem{LLP} Lee T D, Low F E and Pines D  1953 {\it Phys.Rev.} {\bf 90} 297
\bibitem{Feynman}Feynman R P 1955 {\it Phys.Rev.} {\bf 97} 660
\bibitem{Landau} Landau L D and Pekar S I 1946 {\it Zh.Eksp.Teor.Fiz.} {\bf 16} 341
\bibitem{Larsen} Larsen D M 1968 {\it Phys.Rev.} {\bf 172} 967
\bibitem{review1} Mitra T K, Chatterjee A and Mukhopadhyay S 1987 {\it Phys.Rep.} {\bf 153} 91
\bibitem{review2} Alexandrou C and Rosenfelder R 1992 {\it Phys.Rep.} {\bf 215} 1
\bibitem{kelin} Wang K L, Chen Q H and Wan S L 1994 {\it Phys.Lett.A} {\bf 185} 216
\bibitem{yi} Wang K L, Wang Y and Wan S L 1996 {\it Phys.Rev.B} {\bf 54} 12852
\bibitem{Smondyrev} Smondyrev M A 1987 {\it Theor.Phys.} {\bf 68} 653
\bibitem{Selyugin} Selyugin O V and Smondyrev O V 1989 {\it Phys.Stat.Sol.(b)} {\bf 155} 155
\bibitem{exciton} Wang Y, Wang K L and Wan S L 1996 {\it Phys.Rev.B} {\bf 54} 1463
\bibitem{Jiang} Wang K L, Jiang W, Wan S L and Pan H J 2000 {\it Phys.Rev.E} {\bf 61} 4795
\bibitem{Han}  Han R S, Gao X L and Wang K L 2000 {\it Phys.Rev.B} {\bf 62} 15579
\bibitem{Anderson} Wang K L, Gao X L and Lin Ji 1999 {\it Phys.Rev.B} {\bf 60} 15492
\bibitem{1DP} Chen Q H, Wang K L and Wan S L 1994 {\it J.Phys-Condens.Mat.} {\bf 6} 6599
\bibitem{Holstein} Han R S, Lin Z J and Wang K L  2002 {\it Phys.Rev.B} {\bf 65} 174303

\end{thebibliography}
\end{document}